# Calibration-Free Travel Time After Photobleaching Velocimetry

Audrey J. Wang[1], Jianyu Deng[2], David Westbury[2], Yi Wang[2], and Guiren Wang[2,3*]

[1] Chemistry and Biochemistry Department, University of South Carolina
[2] Department of Mechanical Engineering, University of South Carolina
[3] Biomedical Engineering Program, University of South Carolina, Columbia, SC, 29208, USA

* Correspondence: guirenwang@sc.edu

In interfacial science, there is an increasing need to measure flow velocity fields at interfaces with ultrahigh spatial and temporal resolution to study transport phenomena. Although laser-induced fluorescence photobleaching anemometry (LIFPA) has achieved nanoscopic resolution for flow measurement, it requires pre-calibration, which is unavailable for unknown flows. We present a novel, calibration-free travel time after photobleaching velocimeter (TTAPV) which can both measure fluid flow velocity and satisfy the long-anticipated need of calibration for LIFPA.

In many interfacial flows, such as in micro- and nanofluidics, AC electrokinetics, flows inside electric doubles layers, capillary blood flows, and flows in porous media, there is an increasing need for the ability to measure flow velocity profiles. Such a technique that can achieve both ultrahigh spatial and temporal resolution has been one of the major efforts in fluid dynamics. One potential application of this technique is to determine whether flow has a slip or non-slip boundary condition over a solid surface, a topic that has been heavily debated over the past two centuries with no definitive conclusion [1-3]. Another area where flow is poorly understood is in porous media, such as in human tissues, oil shale, groundwater contamination, and geological carbon dioxide storage, where the hydraulic diameter of pores can range from a few micrometers down to tens of nanometers [4,5]. Recently, there has also been a growing interest in the effect of hydrophobicity on fluid-solid interactions [6-8], where slip flow is an important focus of interest.

Current available techniques to measure flows primarily rely on particles as tracers. Some examples include the widely known micro- and nanoparticle image velocimetry (µPIV [9,10] & nanoPIV [11,12]) and laser Doppler velocimetry [13]. While particles have been assumed to reliably represent their surrounding fluids, this is often not the case. In fact, in many microflows, particles do not have the same velocity as their surrounding fluids, such as in electrokinetics, magnetophoresis, acoustophoresis, photophoresis, thermophoresis, and near wall flows [14,15]. Additionally, the best resolution achieved thus far through PIV is on the similar order of or even larger than the measured slip length, which makes it difficult to determine whether or not there is slip flow [16,17]. If the size of the resolution is larger than the slip length, it is impossible to directly measure the slip length accurately. Furthermore, most measured slip lengths are shorter than 100 nm, yet almost all optics-based methods used today suffer from the diffraction limit of roughly 200 nm [18,19], limiting their spatial resolution.

The use of neutral molecular dye as a tracer bypasses the issues of using particles as tracers. However, current methods tend to suffer from relatively limited temporal and spatial resolution (also due to the diffraction limit), such as molecular tagging velocimetry [20] and fluorescence photobleaching recovery [21]. A periodic photobleaching method has been developed that can measure the bulk average flow by measuring the time of migration of a photobleached blot in the flow [22]. This method requires two lenses, one for photobleaching the dye and the other for measuring the fluorescence signal. Thus, it also requires pre-calibration to estimate the distance between the two laser foci and has limited resolution that can only measure bulk flow velocity. Additionally, the use of two lenses severely limits this method from being compatible with a microscope, which has only one objective.

Laser-induced fluorescence photobleaching anemometry (LIFPA) [15,23] is a method which uses small, neutral molecular dye as a flow tracer to measure flow velocities with both high temporal and spatial resolution. As a result, LIFPA has led to several new discoveries recently [24-26]. With the revolutionary technology of stimulated emission depletion (STED) [27], LIFPA, which is compatible with STED, has achieved a spatial resolution of 70 nm [28]. As Cuenca et. al mentioned [29], STED-LIFPA is the only method thus far that has been able to acquire three-dimensional velocity profiles below the microscale. Despite this achievement, as pointed by Schembri et. al [30], the main drawback of this method lies in its need for calibration of light intensity prior to velocity measurements, which is not available for unknown flows. This calls for a calibration-free velocimeter that can be made compatible with LIFPA's setup to help calibrate LIFPA.

To address this need, we present a calibration-free single point travel time after photobleaching velocimeter (TTAPV) that can measure flow velocity profiles in a capillary with high spatial resolution up to the diffraction limit (~ 200 nm). It uses small, neutral molecular dye as a tracer, which can faithfully follow its surrounding fluid, and is compatible with a microscope for ease of use. Additionally, our setup shares the same optical setup with LIFPA, making it readily available for LIFPA calibration at any time if needed. It is also compatible with STED-LIFPA and two-photon absorption-based LIFPA for potential nanosocopic measurements in the future.

Just as LIFPA is similar to hot-wire anemometry, TTAPV is similar to hot-wire anemometry with multiple parallel wires, but non-invasive and with high resolution [31]. The proposed method is illustrated in Fig. 1a. Two laser foci are required, which can be achieved using either two lasers (as in the present work) or one laser with a beam splitter. Both lasers focus within a flow field (capillary used in the present work), through which a fluid containing a small, neutral molecular dye flows. In 1D flow measurements, the two laser foci are aligned to the same $y$- and $z$-positions at the capillary, but different streamwise $x$-positions. The distance between the streamwise foci is the travel distance length $L$.

The upstream laser, referred to as the photobleaching laser, or laser P in Fig. 1a, is positioned at position P' in the capillary. The photobleaching laser is periodically pulsed with a given pulse width $t_{pw}$ and frequency $f_p$. A portion of this light is directed toward a photon detector, allowing pulses to be identified as peaks in a time series of light intensity, indicating the time of photobleaching. With each pulse, the laser photobleaches a portion of the moving dye to generate a photobleached blot of fluid. The second laser, referred to as the detection laser, or laser D, is a continuous wave (cw) laser that focuses downstream at position D'. Fluid inside the capillary flows from position P' to D', and the distance between these positions is equal to $L$. The detection laser is a probe laser that continuously induces fluorescence in the moving dye inside the capillary. The fluorescence intensity $I_f$ at the focus of the detection laser is measured by the same detector that measures the photobleaching pulses. When a blot of bleached dye reaches position D', a dip in $I_f$ is observed.

In the final time series of light intensity, as shown in Fig. 2, the time measured from a peak to its corresponding dip indicates the travel time $t_t$, or the time it takes for the dye to travel a distance of $L$. If the distance $L$ between positions P' and D' can be directly measured by a camera, then the flow velocity $v$ can be determined by the following:

$$v = \frac{L}{t_t} \qquad (1)$$

Since this method has a microscale $L$ and sub-nanoscale laser focus, Taylor dispersion has a negligible effect on the blot's deformation.

Several important aspects must be considered for high resolution measurement. First, although two laser foci are required for their respective functions, both must pass through one objective to be fully compatible with a microscope and to measure $L$ accurately. To do this, rather than having the lasers enter the pupil of the objective parallel to each other, they should enter at an angle, which can be adjusted by tuning mirror 2 shown in Fig. 1a, which changes the angle of the photobleaching laser beam. This allows the detection laser to enter the pupil parallel to the axis of the objective, while the photobleaching laser enters at an adjustable angle. By maintaining the alignment of the detection laser with the axis of the objective, this technique becomes compatible with LIFPA. The angled entrance of the two lasers through one objective also allows for a sufficiently small $L$, which can minimize velocity measurement complications caused by fluorescence recovery due to molecular diffusion at the bleached blot of dye. Since fluorescence recovery increases with increasing $t_t$, the dip in fluorescence signal becomes less pronounced at low velocities. Therefore, $L$ should be adjustable to enable measurement of a large dynamic range of velocities.

Finally, to ensure that this method is indeed calibration-free, $L$ should be directly measurable. To do this, a stage micrometer (not shown here) is placed at the focal point of the objective, and the distance between the two laser foci is measured on the micrometer using a camera. An image from the camera measuring the distance $L$ within the capillary in the present study is shown in Fig. 1b.

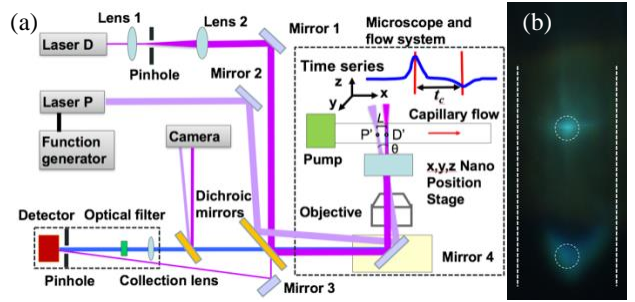

FIG 1. Optical and flow system setup for TTAPV. (a) Schematic of the optical and flow system. (b) Image of the two laser foci at the capillary taken with a camera. The laser focused on the bottom is the photobleaching laser, and the one on top is the detection laser. White dashed lines on the sides represent capillary walls (inner diameter 50 µm).

A schematic of the newly developed TTAPV is illustrated in Fig. 1a. Two laser diodes of wavelength 405 nm were used, a photobleaching laser (laser P) and a detection laser (laser D). The photobleaching laser (250 mW) was a pulsed laser used to generate photobleaching, and the detection laser (7 mW) was a cw laser used to induce fluorescence signal $I_f$ in the moving dye through the capillary. A function generator (Tektronix AFG3102) was applied to the photobleaching laser to generate the pulses, allowing for adjustable pulse width $t_{pw}$ and frequency $f_p$, which were modified based on velocity; at higher velocities, $t_{pw}$ was decreased and $f_p$ was increased, and vice versa.

To achieve the smallest laser focus diameter possible for high spatial resolution, the detection laser was passed through a beam expander. Both lasers were reflected off a dichroic mirror before entering a confocal microscope (Nikon C2). The portion of the photobleaching laser that passed through the dichroic mirror was redirected to a single photon detector (ID100-MMF50 from Photonic Solutions Ltd) with a mirror to monitor periodic photobleaching peaks in $I_f$ time series. The detection laser was directed into the microscope's objective (60X, NA/1.4) at an incident angle of 0 degrees, whereas the photobleaching laser entered at an angle θ. This enabled a tunable travel distance length $L$ between the laser foci at the same $y$- and $z$-positions in the capillary by adjusting θ with mirror 2. $L$ was measured with the use of a stage micrometer at the focal point of the objective and a camera. In the current work, $L$ was set to a fixed distance of 69 µm. Fig. 1b shows the two laser foci within the capillary.

The fluorescence $I_f$ induced by the detection laser at the capillary was collected by the microscope's objective and passed through the dichroic mirror. The signal was then passed through a collection lens, an optical band filter, and a multimode fiber with a core diameter of 10 µm. Finally, the signal was detected by the same detector measuring the pulse signal and recorded onto the hard drive of a computer through an A/D converter (NI USB-6259). The signal was monitored and recorded using LabVIEW. The spatial resolution of our system was dependent on both the diameter of the detection laser at the objective's focal point and the fiber diameter at the detector and was estimated to be about 203 nm based on the diffraction limit.

The present TTAPV method was evaluated using pressure-driven laminar flows through a quartz capillary (inner diameter 50 µm, length 100 mm) driven by a syringe pump (Harvard 11 Pico Plus Elite) through a 10 µL glass syringe (Hamilton). All measurements were performed 50 mm downstream from the entrance of the capillary. Here, fluid flowed in the direction from position P' to D'. All velocity measurements were also taken at the centerline (axis), except for the velocity profile (Fig. 5). We used Coumarin 102 (Sigma-Aldrich Corp., MO) dissolved in water at a concentration of 100 µM as a tracer. This dye has a small molecular size and is electrically neutral, making it an ideal tracer for fluid flow since the dye moves at the same speed as its surrounding fluid and does not interact electrically with the capillary wall or other molecules, unlike larger particles. The capillary was held in place by an XYZ piezo stage (P-545.3C7 PInano® Cap XYZ Piezo Stage from Physik Instrumente), which has a resolution of 1 nm, and was mounted on the confocal microscope. Time series scans of $I_f$ were performed and recorded using PIMikroMove and LabVIEW, and data was processed using MATLAB.

The time series of $I_f$ measured over time for three different centerline velocities $u_c$ is shown in Fig. 2. These time series are a sum of $I_f$ induced by the detection laser and light intensity from laser P's pulse, where the peaks represent each pulse of the photobleaching laser, and the valleys represent the dip in $I_f$ when the bleached blot passes through the focus of the detection laser. The time between the peak and the valley represents the time of travel $t_t$ across the distance $L$.

The time series in Fig. 2 show that when $u_c$ was 0.14 mm/s, 1.42 mm/s, and 28.31 mm/s, the corresponding $t_t$ was 0.5 s, 0.06 s, and 0.002 s respectively. These results demonstrate that $t_t$ decreased as $u_c$ increased, as expected from the Hagen–Poiseuille equation, which confirms that the present TTAPV method can reliably measure flow velocity.

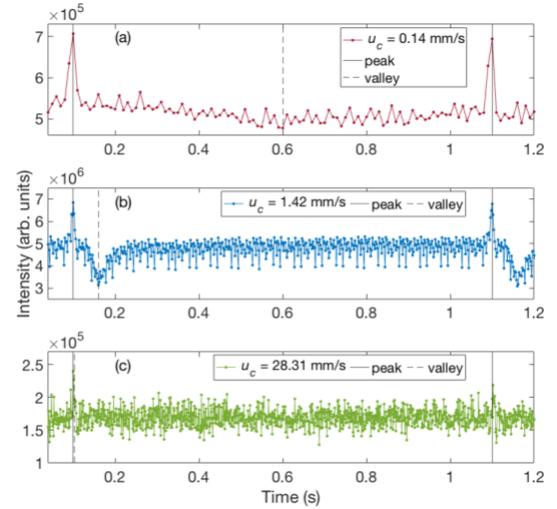

FIG. 2. Time series comparison of different velocities and each of their respective measured travel times. The black solid line, or peak, represents the time of photobleaching, and the black dashed line, or valley, represents the time when the bleached blot reached the detection laser. The time difference between the peak and valley is equal to $t_t$. The pulse frequency was set to 1 Hz for each of these velocity measurements. (a) $u_c$ of 0.14 mm/s with $t_{pw}$ of 30 ms and $f_s$ of 100 samples per second. (b) $u_c$ of 1.42 mm/s with $t_{pw}$ of 30 ms and $f_s$ of 500 samples per second. (c) $u_c$ of 28.31 mm/s with $t_{pw}$ of 20 ms and $f_s$ of 1000 samples per second.

Because $t_t$ decreased as velocity increased, the sampling rate $f_s$ was also increased with increasing velocity to ensure that $t_t$ could still be detected. For the highest velocity, 28.31 mm/s, shown in Fig. 2c, the vertical lines representing the peak and valley are nearly indistinguishable in the time series due to the small $t_t$ and zoomed-out $x$-axis. Notice also that in lower velocities, there was a broadening in width of the valley, as its lower velocity allowed more time for molecular diffusion at the bleached blot. This change in valley width $\Delta W$ can be approximately explained by

$$\Delta W \sim \sqrt{\frac{t_t}{D}} \qquad (2)$$

where $D$ is the diffusivity of the dye.

The proposed TTAPV method shows great potential for measuring velocity over a large dynamic range. Fig. 3 illustrates the dynamic range, which is determined by the maximum and minimum velocities that can be accurately measured using TTAPV with just one objective and a fixed $L$. In capillary flows, there is a positive linear relationship between flow rate $Q$ and centerline velocity $u_c$, which is observed in Fig. 3. Here, we demonstrate that $u_c$ measured using TTAPV aligns well with theoretical $u_c$ (calculated as two times the theoretical bulk velocity). $u_c$ was measured at various $Q$ for a final measured velocity range of 0.1-28.3 mm/s, yielding a dynamic range of approximately 250. Measured velocities had a mean percent error of 4%. While lower velocities were measured, the percent error was found to be higher due to inconsistencies in the syringe pump's performance at low flow rates ($Q < 0.4$ µL/hr).

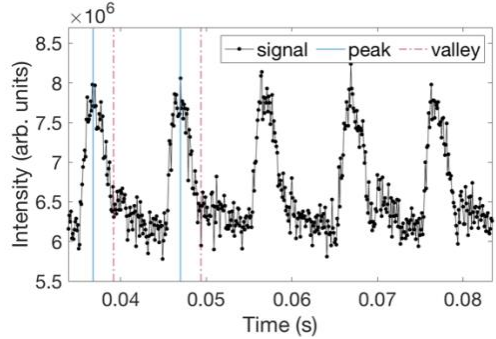

FIG. 4. Temporal resolution demonstrating the smallest measurable length of $t_t$. This $I_f$ data for a velocity of 28.3 mm/s was measured with $t_{pw}$ of 1.2 ms, $f_p$ of 100 Hz, and $f_s$ of 10,000 samples per second.

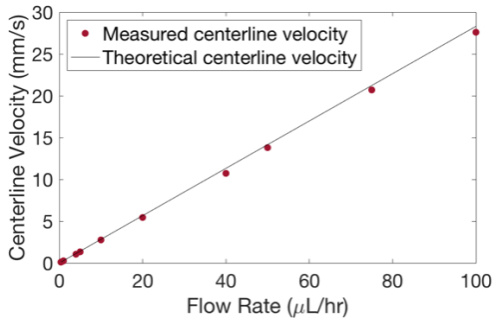

FIG. 3. Dynamic range showing maximum and minimum velocities that can be measured using TTAPV with a fixed $L$ (69 µm). Velocities were measured for a set of flow rates ranging from 0.4-100 µL/hr for a final velocity range of 0.1-28.3 mm/s. Measured velocities had a mean percent error of 4%.

The dynamic range could be increased if needed by adjusting either $L$ or the objective magnification. For example, for measurement of higher velocity, one can increase $L$ or decrease the objective magnification. Determination of the photobleaching pulse width $t_{pw}$ and sampling rate $f_s$ also depends on the velocity being measured. Higher velocities may need a shorter $t_{pw}$ and higher $f_s$ to ensure that the dynamic process can be distinguished and measured.

This technique can measure velocity with high temporal resolution, which is revealed in Fig. 4. Temporal resolution is determined by the smallest measurable length of $t_t$. Fig. 4 shows a time series of $I_f$ for a relatively high $u_c$ of 28.3 mm/s, and demonstrates that this method has a temporal resolution of at least 10 ms, as it can acquire travel time data every 10 ms. Furthermore, since $t_t$ shown in this figure is less than ½ of the time between each pulse, this suggests that our method has the potential for a temporal resolution of 5 ms.

Unlike Fig. 2, the lowest $I_f$ values for each pulse shown in Fig. 4 do not represent the time point where the bleached blot reaches the detection laser; rather, there is a dip in the signal that can be seen before the lowest intensity values immediately following the peak that represents the real $t_t$ of the blot, indicated as the "valley" in this figure. We know this based on the following: (1) this measured $t_t$ is consistent with the theoretical $t_t$ for this velocity; (2) this valley signal is repeatable; and (3) there is no such repeatable valley seen in lower $u_c$. This is likely due to the additional rise and fall time of the photobleaching laser's pulse, which causes the photobleaching signal's fall time to extend past the $t_t$ signal at sufficiently high velocities. The current laser's fall time was measured to be about 2 ms. Thus, to ensure that this method is calibration-free at high velocities, a pulse laser with sufficiently short rise and fall times should be used, so that its total pulse width does not interfere with $t_t$ signal.

TTAPV was also evaluated to measure the velocity profile of pressure-driven laminar flow through the capillary, which is theoretically parabolic. Fig. 5 illustrates both the theoretical and average measured velocity profile of flow (10 uL/hr) in the capillary. This profile was formed by scanning the capillary with 1 µm steps across its diameter and measuring $t_t$ at each radial position. The measured $t_t$ were then used to calculate the velocities at each position with the same $L$. Fig. 5 shows that the measured velocities fit well with the predicted values. Based on the similarity of these profiles, it is clear that TTAPV can measure the velocity profile of flow in a capillary.

It is interesting to note that in the measured velocity profile, the velocities near the wall ceased to follow the theoretical profile starting at approximately 4 µm away from the wall. Rather, the velocities in the near-wall region began to level out at a nonzero value. While the reasoning behind these results is not entirely clear, this velocity data shares similarities with previous findings [32] and suggests the possibility of a slip boundary condition near the wall. Further study is needed to fully explain this observation.

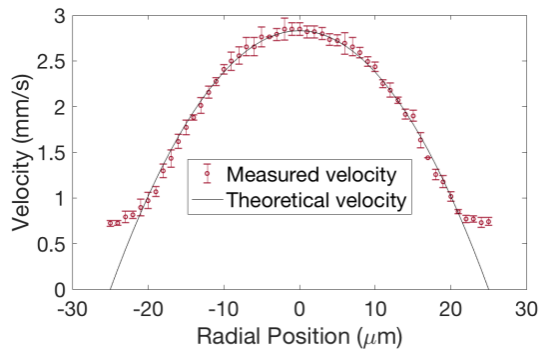

FIG. 5. Velocity profile of fluid flowing at a rate of 10 µL/hr through a capillary with inner diameter of 50 µm. Scans were performed with 1 µm steps in the $x$ (radial) direction with $t_{pw}$ of 25 ms, $f_p$ of 1 Hz, and $f_s$ of 1000 samples per second. Measured data includes error bars with ±1 SD.

In conclusion, we have developed a new, single point travel time after photobleaching velocimeter that directly measures fluid flow velocity. This molecular tracer-based method bypasses issues with using particles as tracers and does not require pre-calibration. Furthermore, this method yields high spatial resolution estimated up to the diffraction limit, and its high temporal resolution and adjustable travel distance allow it to measure a large dynamic range of velocities. In addition to being compatible with a microscope for ease of use, it has further been developed to be compatible with LIFPA, for which it can be used for calibration, raising the potential to achieve a new velocimetry system with ultrahigh spatiotemporal resolution.

While the current development is still limited by the diffraction limit, it has the potential to be integrated with STED-LIFPA, which could greatly enhance the resolution to pass the diffraction limit. Each of these new techniques share similar optical setups that can precisely define the exact position of velocity measurements, especially in the near-wall region, holding great promise for slip flow and other interfacial flow measurements. Ultimately, this new system could change the paradigm of flow velocity profile measurement for studies of transport phenomena in interfacial science.

We thank Akrm Abdalrahman for his help in preparing the flow setup. We also thank Austin Downey and Alexander Vereen for their technical support. This work was supported by the National Science Foundation under Grant No. MRI CBET-1040227.